\documentclass[aps,11pt,nofootinbib,preprintnumbers,showpacs]{revtex4}
\pdfoutput=1
\usepackage{amssymb, amsfonts, amsmath, bm}
\usepackage[pdftex]{graphicx,color}

\begin{document}




\title{
Stable bound orbits around a supersymmetric black lens
}

\author{Shinya Tomizawa$^1$}
\email{tomizawasny@stf.teu.ac.jp}

\author{Takahisa Igata$^{2}$}
\email{igata@rikkyo.ac.jp}
\affiliation{${}^1$ Department of Liberal Arts,~Tokyo University of Technology,~5-23-22, Nishikamata, Otaku, Tokyo 144-8535, Japan\\
${}^2$Department of Physics, Rikkyo University, Toshima, Tokyo 175-8501, Japan\\
}
\preprint{RUP-19-22}




\begin{abstract} 
In higher-dimensional Schwarzschild black hole spacetimes, there are no stable bound orbits of particles. 
In contrast to this, it is shown that there are stable bound orbits in a five-dimensional black lens spacetime.
\end{abstract}

\pacs{04.50.+h  04.70.Bw}
\date{\today}
\maketitle




It is an issue of physical interest whether a black hole spacetime has bound orbits of particles. 
The existence of stable circular orbits is one of characteristic features in the four-dimensional black holes. 
For instance, let us consider the effective potential of the radial motion of a massive particle in Schwarzschild spacetimes. 
For the four dimensions, the centrifugal force and gravitational force make a local minimum of the potential corresponding to a stable circular orbit, 
whereas for the higher dimensions, the effective potential has no local minimum because near the event horizon the gravitational potential becomes more dominated than for four dimensions. 
This leads to the absence from stable circular orbits in higher-dimensional cases in contrast to the four-dimensional case.

\medskip

In an asymptotically flat, stationary and bi-axisymmetric five-dimensional spacetime, the spatial cross section of  an event horizon can have nontrivial topologies of a ring $S^1\times S^2$ and lens spaces $L(p,q)$  as well as a sphere $S^3$~\cite{Cai:2001su,Galloway:2005mf,Hollands:2007aj,Hollands:2010qy}. 
For the spherical topology,  the exact solutions were found in five-dimensional Einstein theory~\cite{Tangherlini:1963bw,Myers:1986un} and five-dimensional minimal supergravity~\cite{Breckenridge:1996is}. 
It was shown that in the vacuum black hole spacetime, there are no stable circular orbits in equatorial planes (rotational axes)~\cite{Frolov:2003en,Page:2006ka,Frolov:2006pe,Cardoso:2008bp}. 
Furthermore, for the supersymmetric black hole spacetime,  the complete analytical solution of the geodesic equations was presented and the properties of massive and massless test particle motion were systematically studied~\cite{Gibbons:1999uv,Diemer:2013fza}. 
For the ring topology, the corresponding solutions were found in both the theories~\cite{Emparan:2001wn,Pomeransky:2006bd,Elvang:2004rt}.  
Remarkably, in contrast to a spherical black hole,  it was explicitly shown that the black ring admits the existence of stable bound orbits~\cite{Igata:2010ye,Igata:2010cd,Igata:2013be}.
For the lens space topologies, the supersymmetric solutions with the horizon topologies of $L(n,1)$ were first constructed in the five-dimensional minimal supergravity~\cite{Kunduri:2014kja,Tomizawa:2016kjh}. 
Because it seems that the black lens cannot admit the separability of the geodesic equations, we cannot show the existence/nonexistence of stable bound orbits by merely solving a one-dimensional potential problem in the radial direction.

\medskip
The three-dimensional lens spaces are mathematically defined as simply, quotients of $S^3$ by ${\mathbb  {Z}}/p$-actions. 
Therefore,  as in a black hole spacetime with the horizon topology of $S^3$, we expect intuitively that there may be no stable bound orbits in a black lens spacetime. 
However, in this paper, contrary to the expectation, we indeed show that there are stable bound orbits around a supersymmetric  black lens. 
The metric of the supersymmetric black lens solution can  be written as a timelike fiber bundle over the multi-centered Gibbons-Hawking space~\cite{Gibbons:1979zt}. 
The essential difference from the supersymmetric black hole with the horizon topology of $S^3$ is the existence of centers outside the horizon, so-called {\it nuts}. 
The existence of the centers significantly changes the motion of particles orbiting around the horizon because the potential diverges by the centrifugal force effect of particles orbiting around them. 
This strongly suggests that there exist local minimums of the effective potential, which admits the existence of stable bound orbits of particles.




In the five-dimensional minimal supergravity, the local metric and gauge potential $1$-form of the supersymmetric black lens solutions take the form~\cite{Kunduri:2014kja,Tomizawa:2016kjh}
\begin{eqnarray}
\label{metric}
ds^2&=&-f^2(dt+\omega)^2+f^{-1}ds_{M}^2,\\
A&=&\frac{\sqrt 3}{2} \left[f(d t+\omega)-\frac KH(d \psi+\chi)-\xi \right]\,,
\end{eqnarray}
where $ds^2_M$ is the Gibbons-Hawking metric:
\begin{eqnarray}
ds^2_M&=&H^{-1}(d\psi+\chi)^2+H(dx^2+dy^2+dz^2), \quad 
\chi=\sum_{i=1}^nh_i\frac{z-z_i}{r_i}\frac{xdy-ydx}{|{\bm r}|^2},\\
H&=&\sum_{i=1}^n\frac{h_i}{r_i}:=\frac{n}{r_1}-\sum_{i= 2}^n\frac{1}{r_i}, \label{Hdef}
\end{eqnarray}
where $r_i:=|{\bm r}-{\bm r_i}|=\sqrt{x^2+y^2+(z-z_i)^2}$ [${\bm r}:=(x,y,z)$, ${\bm r}_i:=(0,0,z_i)$] with constants $z_i$ (we assume $z_1=0<z_2<\cdots<z_n$)  and $H$ is a harmonic function with point sources at ${\bm r}={\bm r}_i$ on ${\mathbb E}^3$. 
The vectors $\partial/\partial t$, $\partial/\partial \psi$ and $\partial/\partial \phi:=x\partial/\partial y-y\partial/\partial x$ are Killing vectors, where  $\partial/\partial\phi$ is the coordinate basis in the standard polar coordinates $(x=r\sin\theta\cos\phi,y=r\sin\theta\sin\phi,z=r\cos\theta)$.  
The other quantities are  written as
\begin{eqnarray}
f^{-1}&=&H^{-1}K^2+L,\\
\omega&=&\omega_\psi(d\psi+\chi)+\hat \omega,\\
\omega_\psi&=&H^{-2}K^3+\frac{3}{2} H^{-1}KL+M, \\
\hat \omega&=&\sum_{j=2}^n\left[\frac{n}{2}k_j^3+\frac{3}{2}(k_1k_j^2-k_jl_1)\right]\frac{r-z_j\cos\theta}{z_{j}r_j}d\phi\notag\\
&&-\sum_{i,j=2(i\not=j)}^n\left(-\frac{1}{2}k_j^3+\frac{3}{2}k_i k_j^2 \right)\frac{r^2-(z_i+z_j)r\cos\theta+z_iz_j}{z_{ji}r_ir_j}d\phi \notag \\
&&-\frac{3}{2}\sum_{i=1}^n\left(-\sum_{j=1}^nk_j h_i+k_i\right)\frac{z-z_i}{r_i}d \phi\notag\\
&&-\sum_{j=2}^n\frac{nk_j^3+3(k_1k_j^2-k_jl_1)}{2z_{j}}d\phi-\sum_{i,j=2(i\not =j)}^n\frac{h_ik_j^3+3k_ik_j^2}{2z_{ji}}d\phi,\\
\xi&=-&\sum_{i=1}^nk_i\frac{z-z_i}{r_i}d \phi,
\end{eqnarray}
where $z_{ji}:=z_j-z_i$ and 
\begin{eqnarray}
M=
-\frac{3}{2}\sum_{i=1}^nk_i+\sum_{i= 2}^n\frac{k_i^3}{2r_i},\quad K=\sum_{i=1}^n\frac{k_i}{r_i},\quad 
L=
1+\frac{l_1}{r_1}+\sum_{i=2}^n\frac{k_i^2}{r_i}.
\end{eqnarray}

In the analysis on stable bound orbits,  
it is more advantageous to work in the coordinate basis vectors $(\partial/\partial\phi_1,\partial/\partial\phi_2)$ of $2\pi$ periodicity, instead of $(\partial/\partial\phi,\partial/\partial\psi)$, where these coordinates are defined by $\phi_1:=(\psi+\phi)/2$ and $\phi_2:=(\psi-\phi)/2$.

\medskip



From the requirements of regularity at ${\bm r}={\bm r}_i\ (i=2,\cdots,n)$ and the absence of closed timelike curves around the horizon and ${\bm r}={\bm r}_i\ (i=2,\ldots,n)$, the parameters $(k_{i\ge 1},l_1,z_{i\ge 2})$ must be subject to
\begin{eqnarray}
1+\frac{1}{z_{i}}(l_1-2k_ik_1-nk_i^2)+\sum_{ j=2(j\not=i)}^n\frac{1}{|z_{ji}|}(k_j-k_i)^2  <0,\label{eq:c1}\\
-\frac{3}{2}\sum_{j=1}^nk_j-\frac{3}{2}k_i+\frac{nk_i^3+3k_1k_i^2-3l_1k_i}{2z_{i}}+\sum_{j=2(j\not =i)}^n\frac{(k_j-k_i)^3}{2|z_{ji}|}=0
\label{eq:c2}
\end{eqnarray}
for each $i=2,\ldots,n$ and 
\begin{eqnarray}
k_1^2+nl_1> 0,\quad 
l_1^2(3k_1^2+4nl_1)>  0.\label{eq:R1R2ineq}
\end{eqnarray}
If the parameters are subject to these constraints, the point ${\bm r}={\bm r}_1(=0)$ corresponds to a null degenerate horizon with the spatial cross section of the lens space $L(n,1)$, whereas the points ${\bm r}={\bm r_i}\ (i=2,\ldots,n)$ denote the coordinate singularities like the origin of the Minkowski spacetime.

The case $n=1$ corresponds to the BMPV black hole, for which the inequality (\ref{eq:c1}) and  the condition~(\ref{eq:c2}) are not imposed. 
The case $n=2$ coincides with the black lens solution with the horizon topology of $L(2,1)$~\cite{Kunduri:2014kja},  in which case Eq.~(\ref{eq:c2})  is reduced to
\begin{eqnarray}
z_2=\frac{k_2 \left(3 k_1 k_2+2 k_2^2-3 l_1\right)}{3 (k_1+2 k_2)}(>0), \label{eq:z2ineq_n=2}
\end{eqnarray}
and the inequalities (\ref{eq:c1}) and (\ref{eq:R1R2ineq})  become
\begin{eqnarray}
 1 + \frac{l_1 - 2 k_2k_1 - 2k_2^2}{z_2}<0, \label{eq:c2ineq_n=2}
\end{eqnarray}
\begin{eqnarray}
l_1^2(3 k_1^2 + 8l_1)>0.\label{eq:R1ineq_n=2}
\end{eqnarray}
The shaded regions in Fig. 1-2 show the parameter region where all of the inequalities (\ref{eq:z2ineq_n=2}), (\ref{eq:c2ineq_n=2}) and (\ref{eq:R1ineq_n=2}) are simultaneously satisfied for the normalized $l_1$ by $l_1=1$ and $l_1=-1$, respectively.

 \begin{figure}[h]
 \begin{tabular}{cc}

\begin{minipage}[t]{0.4\hsize}
 \centering
\includegraphics[width=7cm]{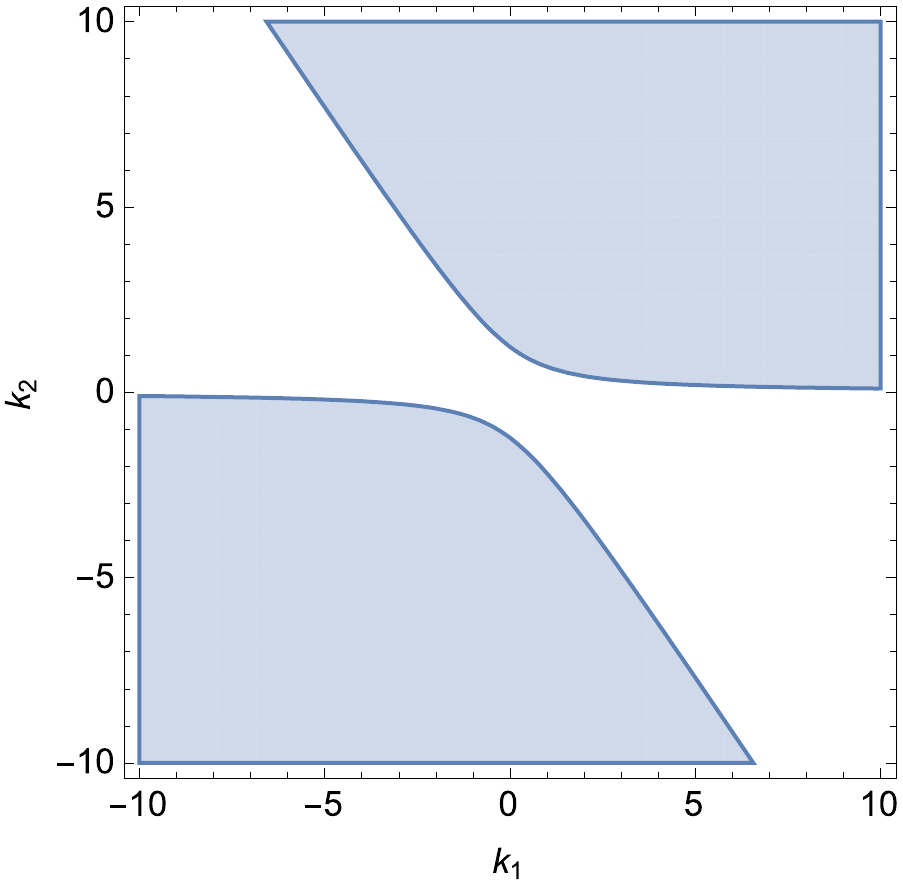}
\caption{$l_1=1$}
 \end{minipage} &\ \ \ \ \ \ \ \ \ \ \ \ \ \ \ \ \ 
 
 \begin{minipage}[t]{0.4\hsize}
 \centering
\includegraphics[width=7cm]{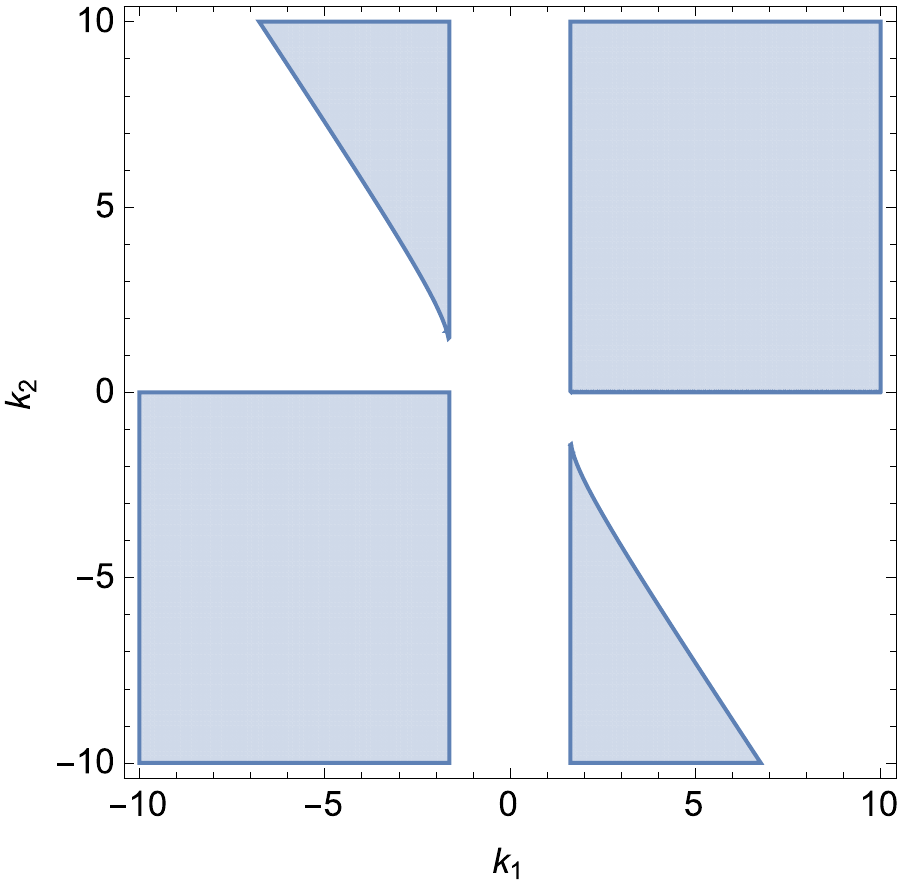}
\caption{$l_1=-1$}
 \end{minipage} 

\end{tabular}
\label{fig:ctc}
\end{figure}






The Hamiltonian of a free particle with the mass $m$ is given by
\begin{eqnarray}
\mathcal{H}=g^{\mu\nu}p_\mu p_\nu+m^2,
\end{eqnarray}
where $p_\mu$ is the canonical momentum. 
It is obvious from independence of ${\mathcal H}$ on the coordinates $(t,\phi_1,\phi_2)$ that $(p_t,p_{\phi_1},p_{\phi_2})$ are constants of motion, which we denote by
 $(p_t,p_{\phi_1},p_{\phi_2})=(-E,L_{\phi_1},L_{\phi_2})$. 
Then, the Hamiltonian can be written in terms of these constants as
\begin{eqnarray}
\mathcal{H}=g^{rr}p_r^2+g^{\theta\theta}p_\theta^2+E^2\left(U+\frac{m^2}{E^2}\right),
\end{eqnarray}
where $U$ is the effective potential defined by
\begin{eqnarray}
U=g^{tt}+g^{\phi_1\phi_1}l_{\phi_1}^2+g^{\phi_2\phi_2}l_{\phi_2}^2-2g^{t\phi_1} l_{\phi_1}-2g^{t\phi_2} l_{\phi_2}+2g^{\phi_1\phi_2}l_{\phi_1} l_{\phi_2},
\end{eqnarray}
with $l_{\phi_1}:=L_{\phi_1}/E$ and $l_{\phi_2}:=L_{\phi_2}/E$. 
The massive and massless particles move on the two-dimensional space $(r,\theta)$ subject to the Hamiltonian constraint ${\mathcal H}=0$. 
In what follows, we consider that the particles (with nonzero energy $E\not=0$) move in the two-dimensional potential $U$, where the allowed regions of the motions for massive and massless particles correspond to $U\le -m^2/E$ and $U\le0$, respectively. 

\medskip
For simplicity, we focus on  the behavior of the effective potential $U$ on the $z$-axis (i.e., $\theta=0,\pi$) of ${\mathbb E}^3$ in the Gibbons-Hawking space. 
The $z$-axis splits up into  the $3$ intervals: 
$I_-=\{(x,y,z)|x=y=0,  z<0\}$, $I_1=\{(x,y,z)|x=y=0,0<z<z_{2}\}$ and $I_+=\{(x,y,z)|x=y=0,z>z_2\}$. 
Now, we elucidate from the shape of the potential the existence of the stable bound orbits of the particles on $I_\pm$ and $I_1$, separately. 

\begin{figure}[h]
 \begin{tabular}{cc}
 \begin{minipage}[t]{0.5\hsize}
\includegraphics[width=8cm]{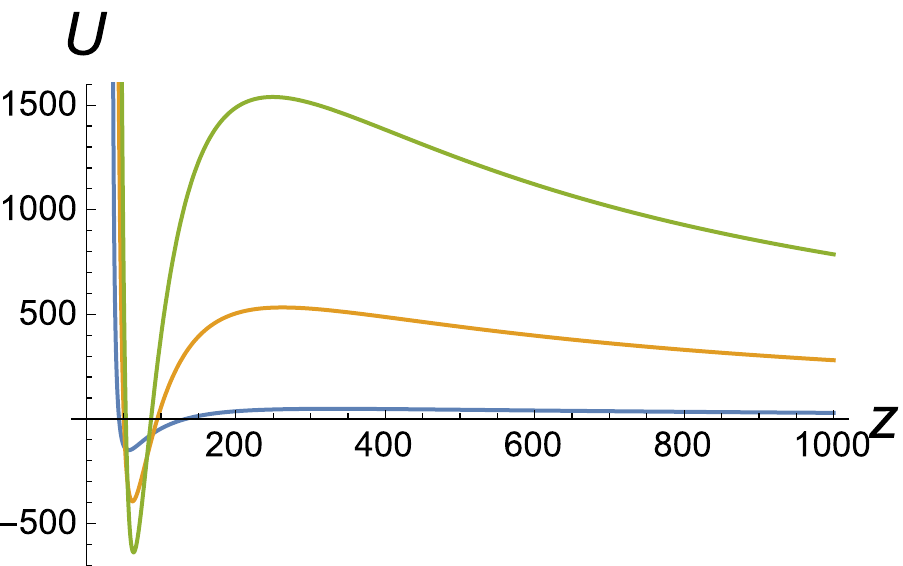}
 \end{minipage} &
 
 \begin{minipage}[t]{0.5\hsize}
\includegraphics[width=8cm]{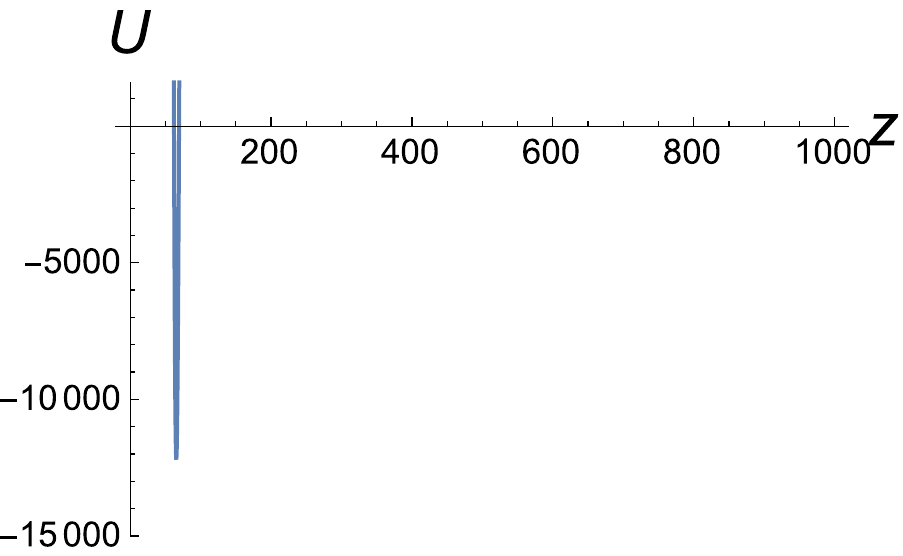}
 \end{minipage}

\end{tabular}
\caption{Effective potentials on  $I_+\ (z>z_2\approx 32.8)$ of the $z$-axis for the black lens with $(k_1,k_2,l_1)=(0,10,1)$.  In the left figure, the  blue, orange and green curves correspond to $(l_{\phi_1},l_{\phi_2})=(-400,0),(-1200,0)$ and $(-2000,0)$, respectively. In the right figure, the curve corresponds to $(l_{\phi_1},l_{\phi_2})=(-40000,0)$.}
\label{fig:potential_on_I_+}\end{figure}

\medskip
First,  let us see the shapes of the effective potential on $I_+$. 
In the left figure of Fig.~\ref{fig:potential_on_I_+},
the  blue, orange and green curves correspond to $(l_{\phi_1},l_{\phi_2})=(-400,0),(-1200,0)$ and $(-2000,0)$, respectively, and in the right figure the curve corresponds to $(l_{\phi_1},l_{\phi_2})=(-40000,0)$, where we set the parameters as $(k_1,k_2,l_1)=(0,10,1)$.
At the center $z=z_2 \ (\approx 32.8)$, the effect of the centrifugal force about the point  makes the potential diverge to $\infty$.  
For large $|l_{\phi_1}|$, the potential always has a negative local minimum, so that there exist stable bound orbits for massive particles.  
Futhermore, because the equation $U=0$ has three roots $z=z_{\mathrm{in}},z_{\mathrm{out}},z_\infty$ $(z_{\mathrm{in}}<z_{\mathrm{out}}<z_\infty)$ for $z>z_2\approx 32.8$ such that $U>0$ for $z<z_{\mathrm{in}}$, $U<0$ for $z_{\mathrm{in}}<z<z_{\mathrm{out}}$, $U>0$ for $z_{\mathrm{out}}<z<z_{\infty}$ and $U<0$ for $z>z_{\infty}$
there are stable bound orbits of the massless particles in the range of $z_{\mathrm in}\le z\le z_{\mathrm out}$.  
It is shown from these figures that as $|l_{\phi_1}|$ is larger, the width $\Delta z:=z_{\mathrm{out}}-z_{\mathrm{in}}$ gets smaller, asymptotically approaches to zero for $l_{\phi_1}\to\infty$. 
This shows that there are stable circular orbits of massless particles  for large $|l_{\phi_1}|$ because $\Delta z \approx 0$ means that such massless particles stay at $z=\mathrm{const.} $

\begin{figure}[h]
\includegraphics[width=8cm]{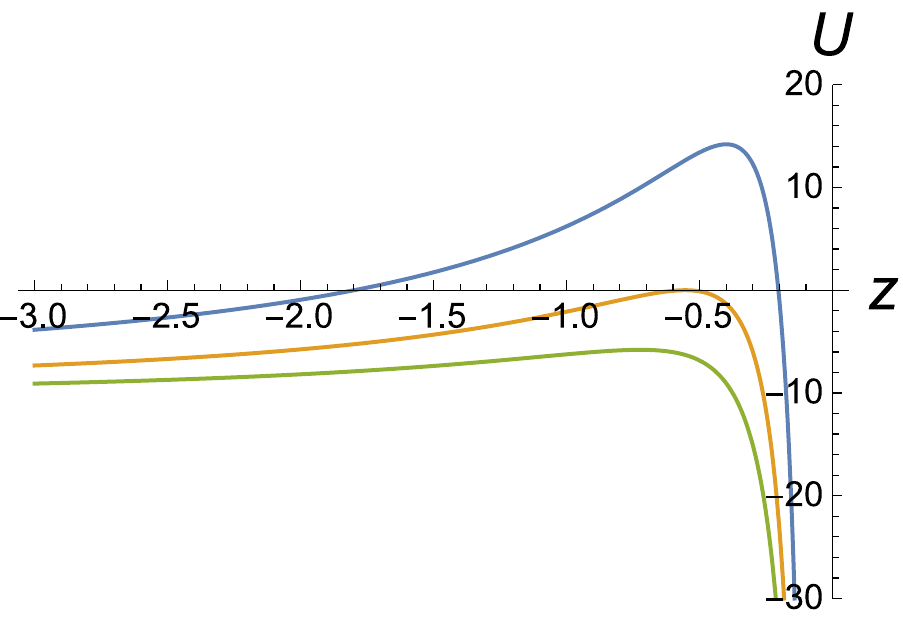}
\caption{Effective potentials on  $I_-\ (z<0)$ of the $z$-axis for the black lens with $(k_1,k_2,l_1)=(0,10,1)$.  The  blue, orange and green  curves correspond to $(l_{\phi_1},l_{\phi_2})= (0,16),(0,13.44...)$ and $(0,12)$, respectively.}
\label{fig:potential_on_I_-}
\end{figure}

\medskip
Next,  we move on the behaviors of the effective potential on $I_-$. 
In Fig.~\ref{fig:potential_on_I_-}, the  blue, orange and green  curves correspond to $(l_{\phi_1},l_{\phi_2})= (0,16),(0,13.44...)$ and $(0,30)$, respectively, in the same choice of the parameters $(k_1,k_2,l_1)$.
At the horizon $z=0$, the gravitational force makes the potential diverge to $-\infty$. 
The potential $U$ seems to have no local minimum but a local maximum, which takes zero for $l_{\phi_1}=0,l_{\phi_2}\approx 13.45$. 
This shows that there exist unbounded orbits of massless particles on $I_-$ (it seems that there exist no stable bound orbits of massive and massless particles as far as we have checked numerically).

 \begin{figure}[h]
 \begin{tabular}{ccc}

\begin{minipage}[t]{0.3\hsize}
\includegraphics[width=5cm]{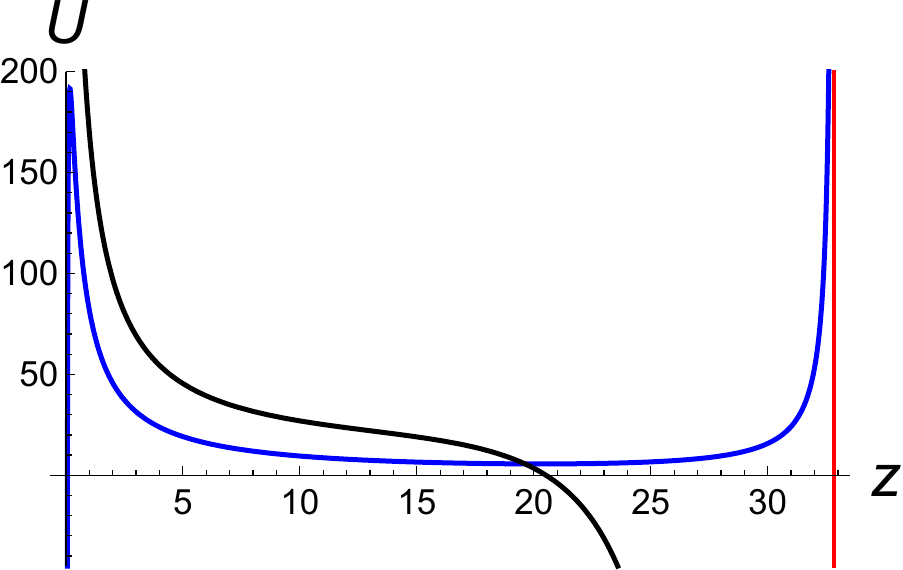}
 \end{minipage} &\ \ 
 
 \begin{minipage}[t]{0.3\hsize}
\includegraphics[width=5cm]{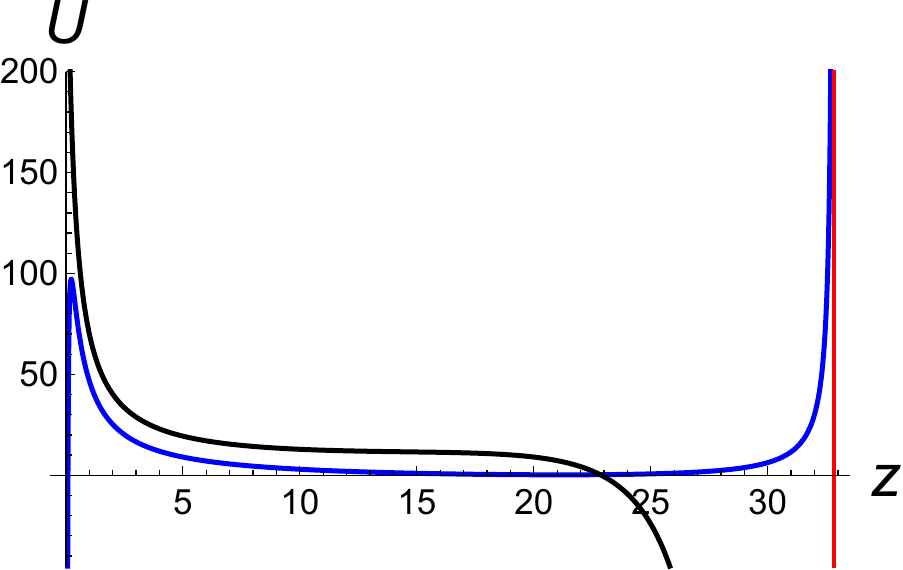}
 \end{minipage} &\ \  
 
 \begin{minipage}[t]{0.3\hsize}
 5\centering
\includegraphics[width=5cm]{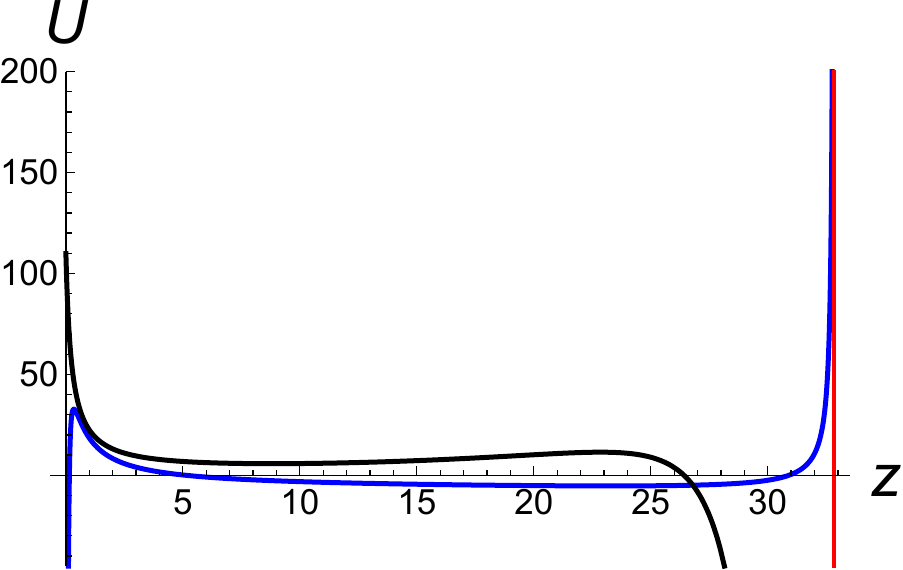}
 \end{minipage} \\

 \begin{minipage}[t]{0.3\hsize}
\includegraphics[width=5cm]{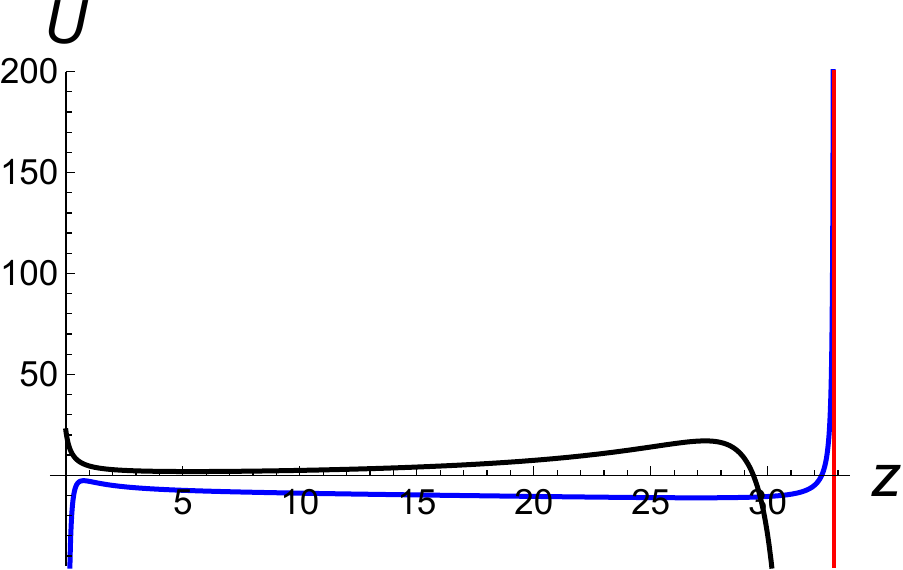}
\end{minipage} &\ \ 
 
\begin{minipage}[t]{0.3\hsize}
\includegraphics[width=5cm]{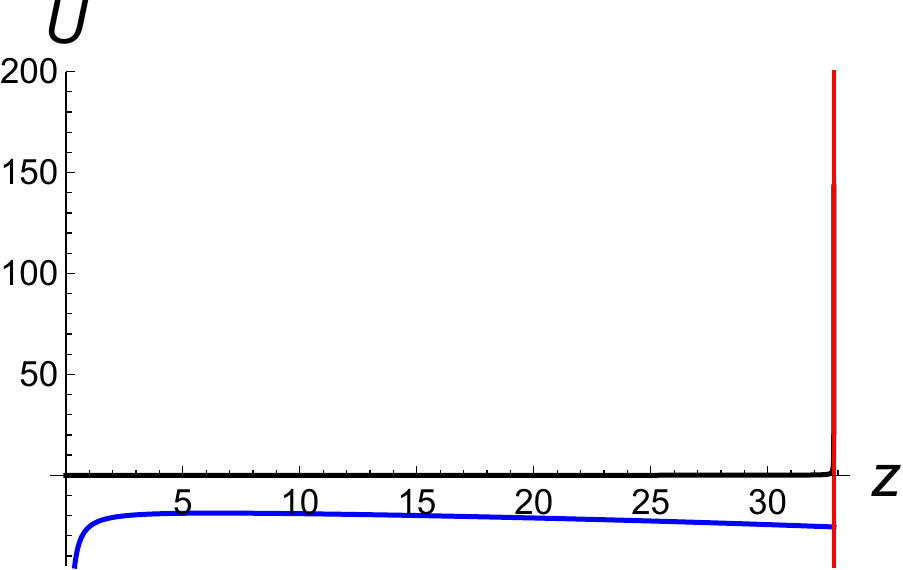}
\end{minipage} &\ \  
 
\begin{minipage}[t]{0.3\hsize}
\includegraphics[width=5cm]{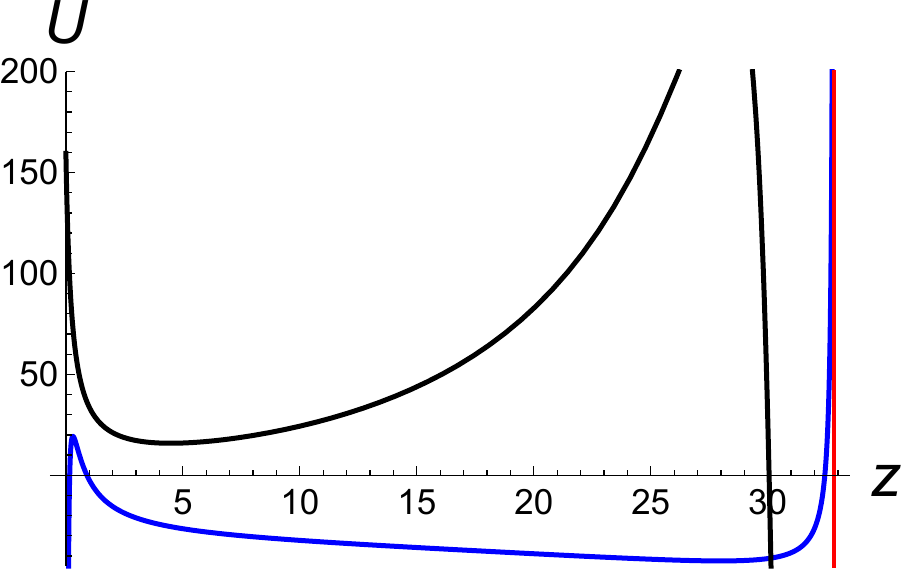}
\end{minipage} \ \

\end{tabular}
\caption{Typical shapes of the potential on $I_1$, where we take the parameters as  $(k_1,k_2,l_1)=(0,10,1)$ and $(l_{\phi_1},l_{\phi_2})=(-4N,2N)$. The upper three graphs correspond to $N=-15,-12,-9,$ from left to right, and the lower three graphs correspond to  $N=-6,0,10$ from left to right.
The endpoints $z=0$ and $z=z_2\approx 32.8$ (red vertical line) of $I_1$ correspond to the horizon and center, respectively.  In each figure, the blue and black graphs denote the potential $U$ and the Hessian divided by $10^{3}$, respectively. }
\label{fig:potential_on_I_1}
\end{figure}

\medskip
Finally, we consider the effective potential on the interval $I_1$, which has the two centers at endpoints $z=z_1(=0)$ and $z=z_2$. 
Note that only the particles with the angular momenta of the special ratio of $L_{\phi_1}/L_{\phi_2}=l_{\phi_1}/l_{\phi_2}=-2$ are allowed to stay on $I_1$ because $I_1$ corresponds to the fixed points of the Killing isometry $v:=\partial/\partial \phi_1+2\partial/\partial\phi_2$, and hence only the particle with a zero angular momentum of $J:=p_\mu v^\mu=L_{\phi_1}+2L_{\phi_2}=0$ can stay on the axis $I_1$
(for $L_{\phi_1}/L_{\phi_2}=l_{\phi_1}/l_{\phi_2}\not=-2$, the potential diverges, which means that $I_1$ corresponds to the potential barrier). 
Figure \ref{fig:potential_on_I_1} shows the typical features of the potential, where we take the same set of parameters $(k_1,k_2,l_1)=(0,10,1)$ and $(l_{\phi_1},l_{\phi_2})=(-4N,2N)$ for $N=-15,-12,-9,-6,0,10$.
In each figure, the blue and black graphs denote the potential and its Hessian, respectively. 
Near the horizon at the center $z=z_1(=0)$, as in the four-dimensional Schwarzschild spacetime, the potential increases by the effect of the centrifugal force, while as closer to the horizon, the strong effect of the gravitational force causes the potential to diverge to $-\infty$.  
On the other hand,  near the other center $z=z_2$, due to the effect of the centrifugal force, 
the potential again increases monotonically and then diverges to $\infty$ except for $N=0\ (l_{\phi_1}=l_{\phi_2}=0)$.  
Thus, the potential also has a positive minimum on $I_1$. 
This is the reason why in contrast to the higher-dimensional Schwarzschild spacetimes, there exist stable bound orbits of massless particles as well as massive particles in a black lens spacetime  with multiple centers.




\label{sec:discuss}
In this paper, we have numerically clarified the existence of stable bound orbits of particles around the supersymmetric black lens with the horizon topology of $L(2,1)$ in the five-dimensional minimal supergravity. 
It can be expected that the stable bound orbits also exist even for the more general topologies $L(n,1)\ (n\ge 3)$ for the following reason. 
The supersymmetric  black lens with the horizon topology $L(n,1)$~\cite{Tomizawa:2016kjh} has $(n-1)$ centers at ${\bm r}={\bm r}_i\ (i=2,\ldots,n)$ outside the horizon, by which the $z$-axis of ${\mathbb E}^3$  in the Gibbons-Hawking space  is split  into  the $(n+1)$ intervals: $I_-=\{(x,y,z)|x=y=0,  z<z_1=0\}$, $I_i=\{(x,y,z)|x=y=0,z_i<z<z_{i+1}\}\ (i=1,...,n-1)$ and $I_+=\{(x,y,z)|x=y=0,z>z_n\}$. 
For the same reason as in the case $L(2,1)$, only the particles with the angular momenta of the ratio $l_{\phi_1}/l_{\phi_2}=-(n-i+1)/(n-i)$  are allowed to stay on $I_i$.
At the centers $z=z_i$ and $z=z_{i+1}$ on $I_i$, the effective potential for particles with such angular momenta diverges to $\infty$ by the centrifugal force effect. 
Therefore, this effective potential is expected to have, at least, a single local minimum  between adjacent two centers in the $z$-direction.
On the other hand, the effective potential for the particles staying on $I_i$ will also make a local minimum in the normal direction to the $z$-axis  (i.e., in the $\theta$ direction) due to the gravitational attractive force of the black lens at $z=z_1(=0)$. 
As a result, we will find stable bound orbits of massive/massless particles in each interval $I_i$.
This may lead to the existence of many stable bound orbits on the $z$-axis. 
We will analyze the more general case in a future paper.

\medskip 
The supersymmetric black lens admits the presence of evanescent ergosurfaces~\cite{Kunduri:2014kja,Tomizawa:2016kjh}, which are defined as timelike hypersurfaces such that a stationary Killing vector field becomes null there and timelike everywhere except there.  
They appear at the points at which $f=0$ corresponding to $H=0$ (for example, for $n=2$ and $k_1=0$, they exist at  $z=2z_2/3$ and $z=2z_2 $ in the $z$-axis).
Reference~\cite{Eperon:2016cdd} proved that on such surfaces, massless particles with zero energy ($E=0$) relative to infinity move along  stable trapped null geodesics.
In the above analysis, we remove such massless particles staying on the evanescent ergosurfaces because 
$(l_{\phi_1},l_{\phi_2})$ are divided by $E$ (hence, for such particles, we must use a different effective potential, for instance, 
$U'=g^{\phi_1\phi_1}L_{\phi_1}^2+g^{\phi_2\phi_2}L_{\phi_2}^2+2g^{\phi_1\phi_2}L_{\phi_1}L_{\phi_2}$). 
However, we can know the motion of particles in the zero energy limit $E\to 0$ from the potential $U$ in the limit $|l_{\phi_1}|\to \infty$ or $|l_{\phi_2}| \to \infty$.
It turns out from the right figure in Fig.~\ref{fig:potential_on_I_+} that for considerably large $|l_{\phi_1}|\ (=40000)$, the potential $U$ has the local minimum at  $z\approx  65.6 $, 
which is just the position of  the evanescent ergosurface on $I_+$
because $2z_2\approx 65.6$. 
This means that there can be stable bound orbits of massless particles with zero energy ($E=0$) staying on the evanescent ergosurface at $z=2z_2$ because $\Delta z \simeq 0$, 
whereas in the above analysis, we have shown that for massless particles with nonzero energy ($E\not=0$), there are also stable bound orbits. 
Furthermore,  it is shown in Ref.~\cite{Eperon:2016cdd} that for the horizonless supersymmetric solutions, the presence of evanescent ergosurfaces makes some linear perturbation decay slowly and leads to nonlinear instability. 
This result does not directly apply to the supersymmetric black lens solution 
but the presence of stable bound orbits of particles with nonzero energy may exhibit  corresponding nonlinear instability.




\acknowledgments
This work was supported by the Grant-in-Aid for Scientific Research (C) [JSPS KAKENHI Grant Number ~17K05452 ~(S.T.) ] and the Grant-in-Aid for Early-Career Scientists [JSPS KAKENHI Grant Number~JP19K14715~(T.I.)] from the Japan Society for the Promotion of Science.




\appendix








\begin{thebibliography}{99}

  
 \bibitem{Hollands:2007aj} 
  S.~Hollands and S.~Yazadjiev,
  ``Uniqueness theorem for 5-dimensional black holes with two axial Killing fields,''
  Commun.\ Math.\ Phys.\  {\bf 283}, 749 (2008)
  [arXiv:0707.2775 [gr-qc]].
  
\bibitem{Hollands:2010qy} 
  S.~Hollands, J.~Holland and A.~Ishibashi,
  ``Further restrictions on the topology of stationary black holes in five dimensions,''
  Annales Henri Poincare {\bf 12}, 279 (2011)
   [arXiv:1002.0490 [gr-qc]].
   
   \bibitem{Cai:2001su} 
  M.~l.~Cai and G.~J.~Galloway,
  ``On the Topology and area of higher dimensional black holes,''
  Class.\ Quant.\ Grav.\  {\bf 18}, 2707 (2001)
  [hep-th/0102149].
 

\bibitem{Galloway:2005mf} 
  G.~J.~Galloway and R.~Schoen,
  ``A Generalization of Hawking's black hole topology theorem to higher dimensions,''
  Commun.\ Math.\ Phys.\  {\bf 266}, 571 (2006)
  [gr-qc/0509107].



\bibitem{Tangherlini:1963bw} 
  F.~R.~Tangherlini,
  ``Schwarzschild field in n dimensions and the dimensionality of space problem,''
  Nuovo Cim.\  {\bf 27}, 636 (1963).
 
  \bibitem{Myers:1986un} 
  R.~C.~Myers and M.~J.~Perry,
  ``Black Holes in Higher Dimensional Space-Times,''
  Annals Phys.\  {\bf 172}, 304 (1986).


  \bibitem{Breckenridge:1996is} 
  J.~C.~Breckenridge, R.~C.~Myers, A.~W.~Peet and C.~Vafa,
  ``D-branes and spinning black holes,''
  Phys.\ Lett.\ B {\bf 391}, 93 (1997)
  [hep-th/9602065].
  
  
\bibitem{Page:2006ka} 
  D.~N.~Page, D.~Kubiznak, M.~Vasudevan and P.~Krtous,
  ``Complete integrability of geodesic motion in general Kerr-NUT-AdS spacetimes,''
  Phys.\ Rev.\ Lett.\  {\bf 98}, 061102 (2007)
  [hep-th/0611083].
    
\bibitem{Frolov:2003en} 
  V.~P.~Frolov and D.~Stojkovic,
  ``Particle and light motion in a space-time of a five-dimensional rotating black hole,''
  Phys.\ Rev.\ D {\bf 68}, 064011 (2003)
  [gr-qc/0301016].
    
\bibitem{Frolov:2006pe} 
  V.~P.~Frolov, P.~Krtous and D.~Kubiznak,
  ``Separability of Hamilton-Jacobi and Klein-Gordon Equations in General Kerr-NUT-AdS Spacetimes,''
  JHEP {\bf 0702}, 005 (2007)
  [hep-th/0611245].
  
  
  \bibitem{Cardoso:2008bp}
  V.~Cardoso, A.~S.~Miranda, E.~Berti, H.~Witek and V.~T.~Zanchin,
  ``Geodesic stability, Lyapunov exponents and quasinormal modes,''
  Phys.\ Rev.\ D {\bf 79}, 064016 (2009) 
  [arXiv:0812.1806 [hep-th]].
  
  
   \bibitem{Gibbons:1999uv} 
  G.~W.~Gibbons and C.~A.~R.~Herdeiro,
  ``Supersymmetric rotating black holes and causality violation,''
  Class.\ Quant.\ Grav.\  {\bf 16}, 3619 (1999)
  [hep-th/9906098].
   
  \bibitem{Gibbons:1999uv} 
  G.~W.~Gibbons and C.~A.~R.~Herdeiro,
  ``Supersymmetric rotating black holes and causality violation,''
  Class.\ Quant.\ Grav.\  {\bf 16}, 3619 (1999)
    [hep-th/9906098].
   
   
   \bibitem{Diemer:2013fza} 
  V.~Diemer and J.~Kunz,
  ``Supersymmetric rotating black hole spacetime tested by geodesics,''
  Phys.\ Rev.\ D {\bf 89}, 084001 (2014)
  [arXiv:1312.6540 [gr-qc]].
 
\bibitem{Emparan:2001wn} 
  R.~Emparan and H.~S.~Reall,
  ``A Rotating black ring solution in five-dimensions,''
  Phys.\ Rev.\ Lett.\  {\bf 88}, 101101 (2002)
  [hep-th/0110260].
 


\bibitem{Pomeransky:2006bd} 
  A.~A.~Pomeransky and R.~A.~Sen'kov,
  ``Black ring with two angular momenta,''
  hep-th/0612005.

 
  \bibitem{Elvang:2004rt} 
  H.~Elvang, R.~Emparan, D.~Mateos and H.~S.~Reall,
  ``A Supersymmetric black ring,''
  Phys.\ Rev.\ Lett.\  {\bf 93}, 211302 (2004)
  [hep-th/0407065].

\bibitem{Igata:2010ye} 
  T.~Igata, H.~Ishihara and Y.~Takamori,
  ``Stable Bound Orbits around Black Rings,''
  Phys.\ Rev.\ D {\bf 82}, 101501 (2010)
  [arXiv:1006.3129 [hep-th]].
     
 
\bibitem{Igata:2010cd}
  T.~Igata, H.~Ishihara and Y.~Takamori,
  ``Chaos in Geodesic Motion around a Black Ring,''
  Phys.\ Rev.\ D {\bf 83}, 047501 (2011)
  [arXiv:1012.5725 [hep-th]].
  
  
\bibitem{Igata:2013be} 
  T.~Igata, H.~Ishihara and Y.~Takamori,
  ``Stable Bound Orbits of Massless Particles around a Black Ring,''
  Phys.\ Rev.\ D {\bf 87}, 104005 (2013)
  [arXiv:1302.0291 [hep-th]].
 


\bibitem{Kunduri:2014kja} 
  H.~K.~Kunduri and J.~Lucietti,
  ``Supersymmetric Black Holes with Lens-Space Topology,''
  Phys.\ Rev.\ Lett.\  {\bf 113}, 211101 (2014)
  [arXiv:1408.6083 [hep-th]].
    
   
 \bibitem{Tomizawa:2016kjh} 
  S.~Tomizawa and M.~Nozawa,
  ``Supersymmetric black lenses in five dimensions,''
  Phys.\ Rev.\ D {\bf 94}, 044037 (2016) [arXiv:1606.06643 [hep-th]].



\bibitem{Gibbons:1979zt} 
  G.~W.~Gibbons and S.~W.~Hawking,
  ``Gravitational Multi-Instantons,''
  Phys.\ Lett.\  {\bf 78B}, 430 (1978).
  
\bibitem{Eperon:2016cdd} 
  F.~C.~Eperon, H.~S.~Reall and J.~E.~Santos,
  ``Instability of supersymmetric microstate geometries,''
  JHEP {\bf 1610}, 031 (2016)
  [arXiv:1607.06828 [hep-th]].
 






  
\end{thebibliography}
\end{document}